\documentclass[twocolumn, tighten]{aastex63}
\pdfoutput=1
\usepackage{natbib}
\usepackage{graphicx}
\usepackage{enumerate}
\usepackage{amssymb, amsmath}
\usepackage{color}
\usepackage{ulem}
\usepackage{upgreek}
\turnoffedit

\defcitealias{Brandt+Draine_2012}{BD12}
\defcitealias{Bruzual+Charlot_2003}{BC03}
\defcitealias{Zubko_2004}{ZDA04}
\defcitealias{Weingartner_2001}{WD01}
\defcitealias{Schlegel+Finkbeiner+Davis_1998}{SFD}
\defcitealias{Gordon1998DetectionOE}{G98}
\defcitealias{Szomoru+Guhathakurta_1998}{SG98}

\begin{document}

\title{An Optical Spectrum of the Diffuse Galactic Light from BOSS and IRIS}

\author[0000-0003-4339-2764]{Blake Chellew}
\affiliation{Department of Physics \& Astronomy, University of California, Irvine, Irvine, CA 92617, USA}

\author[0000-0003-2630-8073]{Timothy D. Brandt}
\affiliation{Department of Physics, University of California, Santa Barbara, Santa Barbara, CA 93106, USA}

\author[0000-0001-7449-4638]{Brandon S.~Hensley}
\affiliation{Department of Astrophysical Sciences, Princeton University, Princeton, NJ 08540, USA}

\author[0000-0002-0846-936X]{Bruce T.~Draine}
\affiliation{Department of Astrophysical Sciences, Princeton University, Princeton, NJ 08540, USA}

\author{Eve Matthaey}
\affiliation{Department of Astrophysical Sciences, Princeton University, Princeton, NJ 08540, USA}

\begin{abstract}
We present a spectrum of the diffuse Galactic light (DGL) between 3700 and 10,000\,\AA, obtained by correlating optical sky intensity with far-infrared dust emission. We use nearly 250,000 blank-sky spectra from BOSS/SDSS-III together with IRIS-reprocessed maps from the IRAS satellite. The larger sample size compared to SDSS-II results in a factor-of-two increase in signal to noise. We combine these data sets with a model for the optical/far-infrared correlation that accounts for self-absorption by dust. The spectral features of the DGL agree remarkably well with features present in stellar spectra. There is evidence for a difference in the DGL continuum between the regions covered by BOSS in the northern and southern Galactic hemisphere. We interpret the difference at red wavelengths as the result of a difference in stellar populations, with mainly old stars in both regions but a higher fraction of young stars in the south. There is also a broad excess in the southern DGL spectrum over the prediction of a simple radiative transfer model, without a clear counterpart in the north. We interpret this excess, centered at $\sim$6500\,\AA, as evidence for luminescence in the form of extended red emission (ERE). The observed strength of the 4000\,\AA{} break indicates that at most $\sim$7\% of the dust-correlated light at 4000\,\AA{} can be due to blue luminescence. Our DGL spectrum provides constraints on dust scattering and luminescence independent of measurements of extinction.
\end{abstract}

\keywords{Interstellar scattering, Dust continuum emission}

\section{Introduction}
\label{sec:intro}

No part of the sky is completely dark. The sky background includes airglow, scattered sunlight and moonlight, scattered light and thermal emission from dust in both the solar system and the Galaxy, and extragalactic emission from sources including unresolved galaxies and the cosmic microwave background. Each of these carries unique information. In this paper we measure the diffuse Galactic light (DGL): optical emission, both starlight and nebular emission from gas, scattered by diffuse interstellar dust. 

The DGL was first detected at low Galactic latitudes as excess flux above the zodiacal light and airglow \citep{Elvey+Roach_1937}. Subsequent studies have observed the DGL from the ground \citep{Elsasser+Haug_1960}, from a sounding rocket \citep{Wolstencroft+Rose_1966}, and with data from satellites including Voyager \citep{Murthy+Henry+Holberg_1991}, PIONEER \citep{Matsuoka+Ienaka+Kawara+etal_2011}, and HST \citep{Kawara+Matsuoka+Sano+etal_2017}.
Measurements of the DGL now span wavelengths from the infrared \citep{Arendt+Odegard+Weiland+etal_1998,Sano+Kawara+Matsuura+etal_2015,Sano+Kawara+Matsuura+etal_2016} to the ultraviolet \citep{Lillie+Witt_1976,Hurwitz+Bowyer+Martin_1991,Seon+Edelstein+Korpela+etal_2010}.

The DGL can constrain properties of the interstellar radiation field \citep{Murthy+Henry_1995,Witt+Friedmann+Sasseen_1997}, as well as properties of interstellar dust including grain size distribution and composition \citep{Mathis_1973,Schiminovich+Friedman+Martin+Morrissey_2001,Sujatha+Shalima+Murthy+etal_2005}.
Unfortunately, it always appears together with other sources of radiation. The DGL component can be disentangled, however, if another tracer of Galactic dust is available.
\citet[][hereafter \citetalias{Brandt+Draine_2012}]{Brandt+Draine_2012} measured the DGL spectrum, tracing optical background emission with blank-sky calibration spectra taken by the second phase of the Sloan Digital Sky Survey \citep[SDSS-II,][]{York+Adelman+Anderson+etal_2000,Abazajian+Adelman-McCarthy+Agueros+etal_2009}.
To trace Galactic dust they used measurements of 100\,$\upmu$m intensity by the InfraRed Astronomy Satellite \citep[IRAS,][]{Neugebauer+Habing+vanDuinen+etal_1984} as processed by \cite{Schlegel+Finkbeiner+Davis_1998}, hereafter \citetalias{Schlegel+Finkbeiner+Davis_1998}.

More data, especially optical spectra, have become available since the work of \citetalias{Brandt+Draine_2012}.
The Baryon Oscillation Spectroscopic Survey \citep[BOSS,][]{2013AJ....145...10D}, part of SDSS-III, has observed nearly 250,000 blank-sky spectra, providing new measurements of the diffuse optical background. This presents an opportunity to improve the precision of the measured DGL spectrum and to place additional constraints on dust and the interstellar radiation field. Other surveys including DESI \citep{desicollaboration2016} and GAMA \citep{GAMA_2011} are similarly expanding the number of measured optical spectra.

In this paper we derive new measurements of the DGL using the combination of BOSS/SDSS-III optical spectra and the IRIS reprocessing of the IRAS sky maps \citep{2006ASPC..357..167M}. Our approach builds on that of \citetalias{Brandt+Draine_2012}, generalizing their correlation model to account for self-absorption by dust. We structure the paper as follows.
In Section~\ref{sec:methodology} we describe the BOSS and IRIS data as well as the linear model of \citetalias{Brandt+Draine_2012} and our update to the model. Section~\ref{sec:results} presents our DGL spectra, investigates the robustness of our model and spatial variation in the spectra, and compares to the results of \citetalias{Brandt+Draine_2012}.
In Section~\ref{sec:starlight} we compare our DGL spectra to the predictions of a radiative transfer model, finding clear signatures of scattered starlight and looking for constraints on stellar populations and dust grain size distribution. In Section~\ref{sec:ere} we present evidence for extended red emission (ERE), namely an excess of intensity at $\sim$6500\,\AA{} over the predictions of a radiative transfer model, and we estimate quantities including integrated ERE intensity, comparing to previous detections in the diffuse ISM. Then we derive an upper bound on luminescence at 4000\,\AA{}. We conclude with Section~\ref{sec:conclusion}.

\section{Data and Methodology} \label{sec:methodology}

In this section we summarize the data we use for optical spectra and far-infrared intensity, then detail our methodology for obtaining the spectrum of the DGL. The optical data are blank-sky spectra from BOSS/SDSS-III while the far-infrared data are the IRIS-reprocessed IRAS maps. Our methodology largely follows \citetalias{Brandt+Draine_2012} but with an important modification to account for self-absorption by dust.

\subsection{BOSS Data}
\label{sec:boss_data}

For optical light intensity, \citetalias{Brandt+Draine_2012} used blank-sky spectra from SDSS-II, the second phase of SDSS.
In this paper we use spectra from BOSS, part of SDSS-III.
The blank-sky spectra are taken for sky calibration. Of the 640 simultaneous spectra per plate in SDSS-II, typically 32 were taken from regions of the sky with no detectable sources. In contrast, the BOSS spectrographs accommodated 1000 simultaneous spectra, and BOSS observations used at least 80 blank-sky spectra per plate. Each BOSS plate has a radius of 1.5\,deg and an area of 7.0\,$\deg^2$.

The blank-sky spectra were reduced in the same way as the science spectra, with the same sky subtraction and spectrophotometric calibration.
Since the calibration was intended for point sources, a correction must be applied when looking at extended sources. \citetalias{Brandt+Draine_2012} found that division by 1.38 is appropriate for SDSS-II, and we apply the same factor. Any difference in this value for BOSS is degenerate \edit1{with the bias} in our DGL spectrum (Section~\ref{sec:bias_factor}) and does not otherwise affect our results.

There are more data in BOSS than SDSS-II; DR12 contains 239,000 blank-sky spectra compared to the 92,000 of DR7. Although BOSS fibers are smaller than SDSS-II fibers ($2''$ diameter instead of $3''$), the combined effect is to reduce the uncertainty in our results. As we show in Section~\ref{sec:results}, our uncertainties are dominated by sky coverage and sampling rather than formal statistical errors, and thus benefit fully from the larger number of spectra in BOSS.

\begin{figure*}
\includegraphics[width=\textwidth]{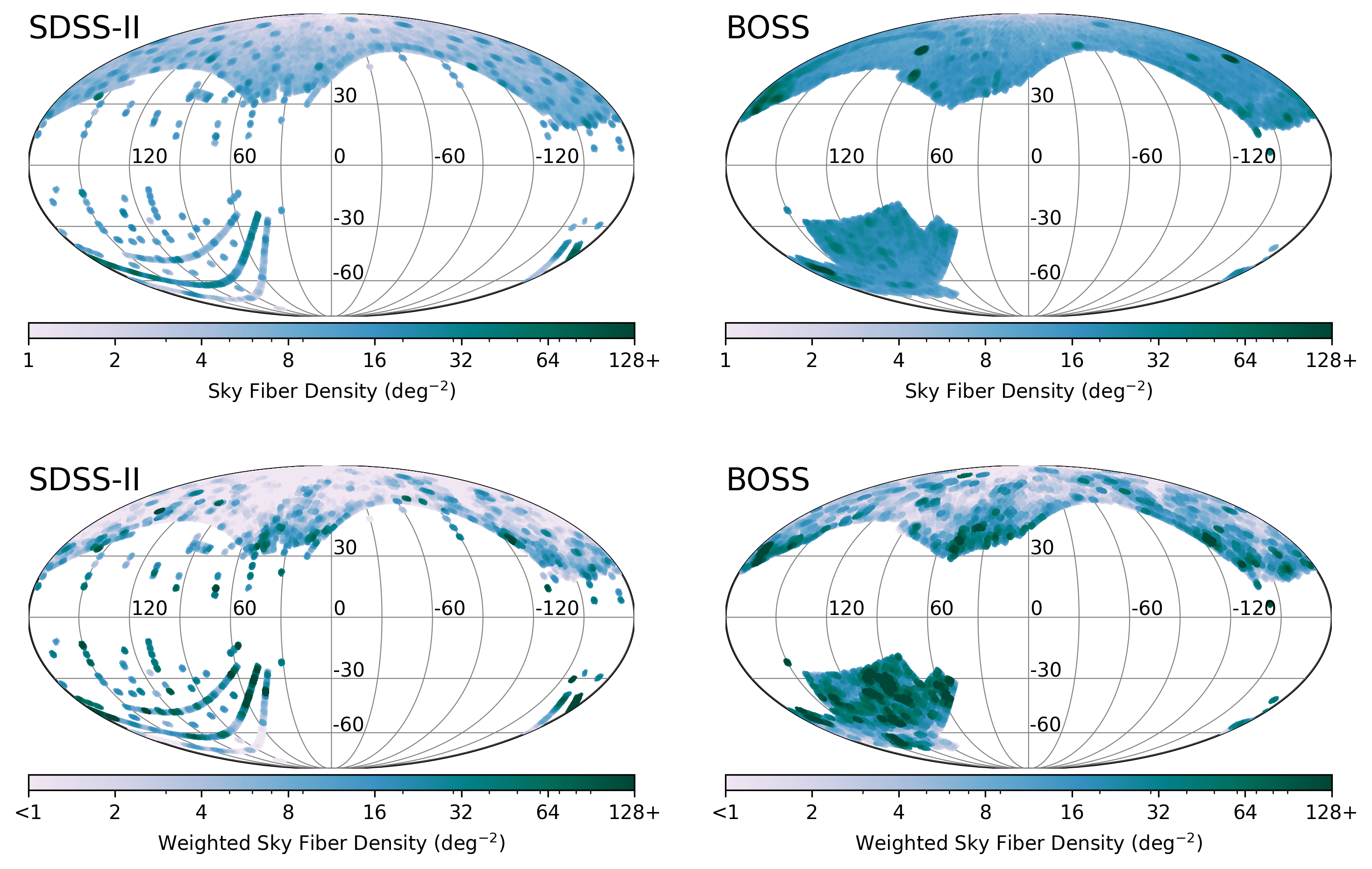}
\caption{Comparison of the sky coverage of SDSS-II (left) and BOSS (right) sky fibers in Galactic coordinates. BOSS has more uniform sky coverage, especially in the southern Galactic hemisphere, with well over twice as many sky fibers overall. The lower panels show fiber density weighted according to variance in $I_{100\,\upmu \rm m}$, which determines a plate's weight in our DGL correlation spectra. The variance is calculated for each plate using the fiber locations, and we multiply the $I_{100\,\upmu \rm m}$ values by a wavelength-dependent optical depth correction factor (Equation~\eqref{eqn:beta}), setting $\lambda = 7000$\,\AA{} for this plot. The handful of plates that we mask due to high $I_{100\,\upmu \rm m}$ or abnormalities in their optical spectra are not included (Section~\ref{sec:boss_data}).}
\label{fig:skymaps}
\end{figure*}

BOSS also has improved sky coverage over SDSS-II, as seen in Figure~\ref{fig:skymaps}. The upper two panels show sky fiber density for SDSS-II and BOSS in Galactic coordinates with $(l, b) = (0, 0)$ at the center. 
BOSS generally has a higher sky fiber density than SDSS-II, with far more coverage in the southern hemisphere resulting in a more uniform patch. The lower two panels show fiber density weighted by the intraplate variance in 100\,$\upmu$m intensity, which determines the impact of fibers on our DGL spectrum. The variance is calculated for each plate using the locations of the sky fibers. From the lower panels we can see that BOSS has more fibers that contribute significantly to the spectrum.

BOSS covers a wider wavelength range than SDSS-II: 3600 to 10,400\,\AA{} rather than 3800 to 9200\,\AA. This means the BOSS uncertainties are much lower for wavelengths near the edges of the SDSS-II range.
It also places the [{\sc O\,ii}]\,$\lambda\lambda$3727-30\,\AA{} nebular emission doublet within the BOSS wavelength range.
The resolving power of the BOSS spectrograph ranges from $R=1560$ to $R=2270$ for the blue channel and from $R=1850$ to $R=2650$ for the red channel.
The wavelengths observed by SDSS-II varied slightly from night to night, so \citetalias{Brandt+Draine_2012} interpolated the spectra onto a common wavelength array. For BOSS, the same set of wavelengths is extracted for every spectrum, so no interpolation is necessary.

We mask all measurements from six fibers that were identified in DR12 as being affected by intermittently bad CCD columns\footnote{https://www.sdss.org/dr12/spectro/caveats}.
We also perform an automated check for plates with an outsized effect on our results, looking at the impact of each plate on the average value of the DGL spectrum over 500\,\AA{} intervals. Then, by visual inspection, we identify twelve plates with a nontrivial effect on the spectrum. We mask nine of these plates\footnote{plates 3666, 4863, 5865, 5371, 6262, 6736, 7253, 7256, and 7261} after finding abnormalities in their optical spectra which appear to be caused by corrupted data and issues with sky subtraction. We do not mask the other three plates, but their impact is not large enough to affect our results.

\subsection{100\,$\mu$m Maps}
\label{sec:100_micron}

\citetalias{Brandt+Draine_2012} used the \citetalias{Schlegel+Finkbeiner+Davis_1998} dust maps to trace far-infrared intensity; these are based on data from the IRAS and COBE \citep{Boggess+Mather+Weiss+etal_1992} satellites.
We instead use the IRIS maps \citep{2006ASPC..357..167M} obtained by reprocessing the IRAS and COBE images with improvements in zodiacal light subtraction, calibration, and destriping. IRIS keeps the $\sim$4$.\!\!^\prime3$ resolution of IRAS.
Our results are consistent with what we find using the \citetalias{Schlegel+Finkbeiner+Davis_1998} maps, and the uncertainties in our results decrease when the IRIS maps are used.

In Section~\ref{sec:updated_model} we introduce a new model for the relationship between optical and far-infrared intensity. This model requires the dust optical depth as a function of position and wavelength, which we obtain from the \citetalias{Schlegel+Finkbeiner+Davis_1998} maps. \citetalias{Schlegel+Finkbeiner+Davis_1998} performed additional processing to convert the 100\,$\upmu$m map to a map of extinction, including removing point sources and estimating the dust temperature. We convert from the \citetalias{Schlegel+Finkbeiner+Davis_1998} $E(B-V)$ values
to wavelength-dependent optical depth using the $R_V = 3.1$ dust model of \cite{Fitzpatrick_1999}.

\subsection{Linear Model from BD12} 
\label{sec:linearmodel}

\citetalias{Brandt+Draine_2012} used a linear model to correlate sky fiber intensity with 100\,$\upmu$m intensity.
Both quantities have a similar dependence on dust column density and dust temperature, as described in Section~2.2 of \citetalias{Brandt+Draine_2012}.
Briefly, scattered intensity is proportional to dust column density times the intensity of illuminating starlight, and this product is proportional to $\tau_{100\,\upmu \textrm{m}}T^{5.5}_{\textrm{dust}}$ if far-infrared emissivity $\propto \nu^{1.5}$ as estimated by \citet{planck_dust_2018}. 
Assuming thermal equilibrium and converting total emission to $I_{100\,\upmu \textrm{m}}$, we also have roughly $I_{100\,\upmu\textrm{m}} \propto \tau_{100\,\upmu \textrm{m}}T^{8}_{\textrm{dust}}$ for dust with $T\approx18$K.
\edit1{If intraplate variations in DGL and $I_{100\,\upmu \textrm{m}}$ are both primarily due to variations in dust column density, rather than variations in illuminating starlight, temperature variations will be small}
\citep{planck_intermediate_2016}. \edit1{Then}
the difference in temperature dependence, $T^{5.5}$ versus $T^8$, \edit1{can be neglected and 
the intraplate variations in DGL can be approximated as linearly dependent on $I_{100\,\upmu \textrm{m}}$.}

The intensities from the SDSS pipeline, $I_{\lambda,\textrm{sky}}$, have already been sky subtracted on a plate-by-plate basis. For consistency the mean value of $(\nu I_{\nu})_{100\,\upmu \textrm{m}}$, the far-infrared intensity, must also be subtracted from each plate. The model is then
\begin{equation}
\lambda I_{\lambda,\textrm{sky},j,p} = \alpha_\lambda \left[(\nu I_{\nu})_{100\,\upmu \textrm{m},j,p} - \langle(\nu I_{\nu})_{100\,\upmu \textrm{m}}\rangle_p\right]
\end{equation}
for fiber $j$, plate $p$, and wavelength $\lambda$, where $\langle\rangle_p$ represents an average across the fiber locations on plate $p$. \citetalias{Brandt+Draine_2012} defines the quantity
\begin{equation}
\label{eqn:x}
x_{j,p} = (\nu I_{\nu})_{100\,\upmu \textrm{m},j,p} - \langle(\nu I_{\nu})_{100\,\upmu \textrm{m}}\rangle_p.
\end{equation} 

The correlation coefficient $\alpha_\lambda$ and its associated uncertainty are calculated using the maximum likelihood estimator (Equations~(5) and (6) of \citetalias{Brandt+Draine_2012}). We use the term ``correlation spectrum'' to refer to the correlation coefficient as a function of wavelength.

This linear model is only accurate in the optically thin limit. As optical depth increases, so does the fraction of DGL photons that are absorbed by dust along our line of sight. \citetalias{Brandt+Draine_2012} therefore masked fibers and plates with high 100\,$\upmu$m intensity, which in turn implies a high dust column density. They applied a threshold of $I_{100\,\upmu \rm m} < 10$ MJy$\,$sr$^{-1}$, corresponding to an optical depth at 5000\,\AA{} of $\tau \approx 0.7$ for 18\,K dust or $\tau \approx 1.1$ for 17\,K dust \citep{2020ApJ...895...38H}.

\subsection{Updated Model to Account for Self-Absorption}
\label{sec:updated_model}

Here we generalize the linear model of \citetalias{Brandt+Draine_2012} to treat self-absorption along the line of sight. This results in a model that is applicable along sight lines with moderate optical depth. We introduce a correction factor,
\begin{equation}
\label{eqn:beta}
\beta_\lambda = \frac{1-\exp(-\tau_\lambda)}{\tau_\lambda},
\end{equation}
updating the definition of $x$ from Equation \eqref{eqn:x} as follows:
\begin{equation}
\label{eqn:x_update}
x_{\lambda,j,p} = ((\nu I_{\nu})_{100\,\upmu m} \beta_{\lambda})_{j,p} - \langle(\nu I_{\nu})_{100\,\upmu m} \beta_\lambda\rangle_p.
\end{equation}
We derive the optical depth $\tau_\lambda$ from 100\,$\upmu$m maps, as described in Section~\ref{sec:100_micron}. Since $\tau_\lambda$ is a function of wavelength, $x$ becomes a function of both wavelength and sky location, while in \citetalias{Brandt+Draine_2012} it was a function only of sky location.

Our model is then
\begin{equation}
\label{eqn:full_model}
\lambda I_{\lambda,\textrm{sky},j,p} = \alpha_\lambda x_{\lambda,j,p},
\end{equation}
and $\alpha_\lambda$ can be found with the same equations as for the linear model except that $x_{j,p}$ is now replaced with $x_{\lambda,j,p}$.
In the optically thin limit ($\tau_\lambda \to 0$) the correction factor becomes $\beta_\lambda = 1$, and we recover the linear model of \citetalias{Brandt+Draine_2012}.
In the optically thick limit ($\tau_\lambda \to \infty$) the correction factor goes to $\beta_\lambda = \tau_\lambda^{-1}$. 
As stated \edit1{in Section~\ref{sec:linearmodel}}, 100\,$\upmu$m emission is roughly proportional to $\tau_{100\,\upmu\textrm{m}}T^{8}_{\textrm{dust}}$, and \edit1{if temperature variations are small enough,} $I_{100\,\upmu \textrm{m}} \propto \tau_{100\,\upmu \textrm{m}}$ is a good approximation. 
Then there are two factors of $\tau$ in Equation~\eqref{eqn:x_update} which cancel so that $x_\lambda$ approaches a constant value for a given wavelength. From Equation~\eqref{eqn:full_model}, $\lambda I_{\lambda,\textrm{sky}}$ also approaches a constant in this limit; only dust within the last $\tau \approx 1$ contributes to the visible scattered light. We expect the use of Equation \eqref{eqn:x_update} over \eqref{eqn:x} to have the largest impact at blue wavelengths, where dust optical depth is highest.

To interpret $\alpha_\lambda$ we consider the model definition of Equation~\eqref{eqn:full_model}. Both $\lambda I_{\lambda,\textrm{sky}}$ and $x_{\lambda}$ are mean-subtracted quantities, but the proportionality constant $\alpha_{\lambda}$ is the same if the means are not subtracted, and so we can write (in an averaged sense):
\begin{equation}
\label{eqn:alpha_corr}
\alpha_\lambda = \frac{\lambda I_{\lambda}}{\left(\nu I_{\nu}\right)_{100\,\upmu \textrm{m}}\beta_\lambda}.
\end{equation}
Because a factor of $\beta_\lambda$ appears here, $\alpha_\lambda$ is no longer proportional to $\lambda I_{\lambda}$. We instead choose to plot 
\begin{equation}
\label{eqn:alpha_prime}
\alpha_\lambda^\prime = \alpha_\lambda \beta_\lambda,
\end{equation}
which has a straightforward interpretation as intensity and is therefore better for comparing to model spectra.

We employ the same masking as \citetalias{Brandt+Draine_2012}, with a threshold of 10\,MJy\,sr$^{-1}$ except where indicated otherwise. We mask based on \citetalias{Schlegel+Finkbeiner+Davis_1998} 100\,$\upmu$m values for the sake of comparison; the IRIS values are generally larger, so if we were to use them for masking we would mask more fibers. The details of masking should not impact our results because, as we will show in Section~\ref{sec:updated_model}, the updated model is insensitive to the masking threshold.

For most analyses we increase the signal-to-noise ratio by binning the correlation spectra with a bin width of 50\,\AA, and we mask each nebular emission line listed in Table~\ref{tab:widths} by removing the data in a region around the central wavelength corresponding to a velocity range of $\pm$200\,km\,s$^{-1}$.

\begin{deluxetable}{cccc}
\label{tab:widths}
\tablewidth{0pt}
\tablecaption{Equivalent widths of nebular emission lines}
\tablehead{
Line (\AA) & \multicolumn{3}{c}{Equivalent Width\tablenotemark{a} (\AA)} \\
& Full Sky & North & South
}
\startdata
[{\sc O\,ii}]~$\lambda\lambda$3727-30 & 16 $\pm$ 5 & 24 $\pm$ 9 & 8 $\pm$ 6 \\
H$\beta$~$\lambda$4863 & 3.9 $\pm$ 0.3 & 4.5 $\pm$ 0.6 & 3.5 $\pm$ 0.4 \\
\phantom{}[{\sc O\,iii}]~$\lambda$4960 & 0.0 $\pm$ 0.2 & 0.1 $\pm$ 0.4 & 0.0 $\pm$ 0.3 \\
\phantom{}[{\sc O\,iii}]~$\lambda$5008 & 1.0 $\pm$ 0.3 & 1.3 $\pm$ 0.6 & 0.8 $\pm$ 0.3 \\
He{\sc \,i}~$\lambda$5877 & 0.2 $\pm$ 0.3 & 0.3 $\pm$ 0.5 & 0.2 $\pm$ 0.3 \\
\phantom{}[{\sc N\,ii}]~$\lambda$6550 & 1.4 $\pm$ 0.3 & 2.9 $\pm$ 0.5 & 0.7 $\pm$ 0.3 \\
H$\alpha$~$\lambda$6565 & 10.7 $\pm$ 0.5 & 15 $\pm$ 1 & 8.7 $\pm$ 0.5 \\
\phantom{}[{\sc N\,ii}]~$\lambda$6585 & 5.6 $\pm$ 0.4 & 9 $\pm$ 1 & 3.7 $\pm$ 0.4 \\
\phantom{}[{\sc S\,ii}]~$\lambda$6718 & 4.6 $\pm$ 0.4 & 8.2 $\pm$ 0.8 & 2.8 $\pm$ 0.4 \\
\phantom{}[{\sc S\,ii}]~$\lambda$6733 & 3.3 $\pm$ 0.3 & 5.6 $\pm$ 0.6 & 2.2 $\pm$ 0.3
\enddata
\tablenotetext{a}{Calculated as described in Section~\ref{sec:lines}; uncertainties are from bootstrap resampling (Section~\ref{sec:bootstrapping})}
\end{deluxetable}

\subsection{Bootstrapping}
\label{sec:bootstrapping}

\citetalias{Brandt+Draine_2012} obtained uncertainties on their DGL spectrum using the errors reported by the SDSS pipeline and the formal uncertainties of a linear fit. This will underestimate the true uncertainty if there are any correlated or non-Gaussian errors, and the effect is amplified when data are binned.
To overcome this limitation we use bootstrap resampling (``bootstrapping'') to compute the uncertainties on our DGL spectra.

Bootstrapping is a technique where many samples of size $N$ are taken from a data set of size $N$, with replacement. This approximates sampling from the true underlying distribution responsible for the data. A statistic is computed for each sample and the variation in the statistic across samples can be used to compute uncertainties.

We use bootstrapping to generate 10,000 realizations of the sample of BOSS sky spectra. Each of these realizations draws from the sample of observed plates rather than fibers. We treat separate observations from the same physical plate (on different nights) as distinct. In this way we approximate repeating the survey many times, each time with a different placement of the BOSS plates. Then we compute a correlation spectrum for each of the 10,000 bootstrap samples and find the 16\textsuperscript{th} and 84\textsuperscript{th} percentiles of $\alpha_\lambda$ at each wavelength.

With the exception of Figure~\ref{fig:compare_errs}, where we also include formal uncertainties, errors shown in figures are half of the 68\% confidence intervals derived from bootstrapping. This directly corresponds to the standard deviation if the errors are Gaussian. In some figures we plot the 16\textsuperscript{th} and 84\textsuperscript{th} percentiles to form an envelope. For binned plots we first bin the correlation spectra before taking percentiles.

\subsection{Bias Factor}
\label{sec:bias_factor}

\edit1{Based on a comparison to the predicted results from stellar models and a radiative transfer model, it appears that $\alpha_\lambda$ is biased low by close to a factor of two.
In Section~\ref{sec:fitting} we describe this comparison, and} we adopt \edit1{bias factors of} $C=2.1$ for the northern Galactic hemisphere and $C=1.9$ for the southern hemisphere.

\edit1{\citetalias{Brandt+Draine_2012} estimated a 20\% uncertainty in the bias factor, based primarily on how the predicted DGL spectrum changes when dust optical depth is varied in the radiative transfer model and when a model of the local ISRF is used instead of stellar population synthesis models.
We adopt the same 20\% uncertainty, resulting in $C = 2.1 \pm 0.4$ in the north and $C = 1.9 \pm 0.4$ in the south.}
There is a 13.5\% gain uncertainty in the IRIS \edit1{100\,$\upmu$m} map \citep{2006ASPC..357..167M}, \edit1{and while this affects the scaling of the measured DGL spectrum, it does not change the percent uncertainty in the bias factor.
Any} error in our flux calibration factor (Section~\ref{sec:boss_data}) \edit1{would have} a similar effect.

\edit1{An estimate of the bias factor should be applied whenever the true ratio of $\lambda I_\lambda$ to $(\nu I_{\nu})_{100\,\upmu \textrm{m}}$ is desired.
We apply bias factors for the} ERE calculations \edit1{in} Section~\ref{sec:num_est}\edit1{, where we are concerned with finding actual intensity values and comparing to previous work.
Because of the high uncertainty in the bias factor,} we do not apply \edit1{it} in plots except where explicitly indicated by a factor of $C$\edit1{.}

\edit1{The source of the bias is not clear. Here we investigate two possibilities: the effect of neglecting 100\,$\upmu$m noise in our model and the difference in resolution between the optical and 100\,$\upmu$m data. We find that neither effect is large enough to explain the observed bias.}

\edit1{Section~4.2 of \citetalias{Brandt+Draine_2012} describes how uncertainties in the 100\,$\upmu$m map,} which are not considered in \edit1{our} model, \edit1{can lead to a bias. However, we estimate the size of this effect and find that the statistical uncertainties in the IRIS map are too small to explain a factor-of-two bias.}
Equation~(14) of \citetalias{Brandt+Draine_2012} \edit1{provides a way to estimate the bias if the measurement uncertainty and the variance in 100\,$\upmu$m intensity are both known.} While the IRIS maps do not include \edit1{uncertainty maps}, \cite{2006ASPC..357..167M} estimates the median noise level to be $0.06 \pm 0.02$\,MJy\,sr$^{-1}$\edit1{. 
We also} find that the median \edit1{variance in 100\,$\upmu$m intensity} across the BOSS plates is 0.06\,MJy$^2$\,sr$^{-2}$. The predicted bias factor is then \edit1{$1.07 \pm 0.04$}.

\edit1{Another possible source of bias is the difference in size between the BOSS fibers
and the IRAS pixels.  
The $\sim$4$.\!\!^\prime3$ IRAS pixels are much larger than the $2''$ BOSS fibers, so there is some error in determining the 100\,$\upmu$m intensity at the location of a BOSS fiber. 
However, we apply Equation~(14) of \citetalias{Brandt+Draine_2012} to find that the bias from this effect is also small.
Projected dust emission closely follows a $P(k) \sim k^{-2.9}$ power law down to 15$''$ \citep{2016A&A...593A...4M}, which means that the power on small scales falls off sharply. 
{\sc H\,i} observations suggest that this $k^{-2.9}$ power law continues down to 5$''$ \citep{2010MNRAS.404L..45R}.
We integrate $P(k)k\,dk$ to estimate the contribution of different spatial scales to the variance.
Assuming that the 
$k^{-2.9}$ power law extends down to 2$''$,
scales from 4$.\!\!^\prime3$ to $3^\circ$ contribute about 25 times as much variance as scales between $2''$ and 4$.\!\!^\prime3$; this alone would result in a bias factor $\lesssim1.05$.}

\edit1{Since random errors in the 100\,$\upmu$m map and the resolution mismatch both predict bias factors much smaller than two, something else must be responsible for the bias. Any deviations from our simple model relating optical intensity to 100\,$\upmu$m emission will contribute to the bias, and we suspect this is the primary cause.
These deviations include physical sources of spatial variation
such as contamination by zodiacal light and associated 100\,$\upmu$m emission, variations in starlight intensity which affect dust grain temperature and therefore 100\,$\upmu$m emission, and variations in the anisotropy or spectrum of incident starlight.
In addition, there might be variations in the physical properties of dust grains arising from processes such as grain growth.}

\citetalias{Brandt+Draine_2012} demonstrated that the bias factor is independent of wavelength for a linear model. Our nonlinear model introduces a possible wavelength dependence, but the nonlinearity correction is $\lesssim20\%$ at all wavelengths (Section~\ref{sec:updated_results}). 

\section{Initial Results}
\label{sec:results}

In this section we present DGL correlation spectra computed using the approach of Section~\ref{sec:updated_model}. We describe the differences between our results and those of \citetalias{Brandt+Draine_2012} due to new data and an updated model. Then we show that our results are insensitive to the masking threshold for plates and fibers with high dust optical depth, and we explore the sensitivity of our results to the spatial footprint of the survey. Finally, we measure the strengths of nebular emission lines.

\subsection{BOSS Results}

\begin{figure*}
\includegraphics[width=\textwidth]{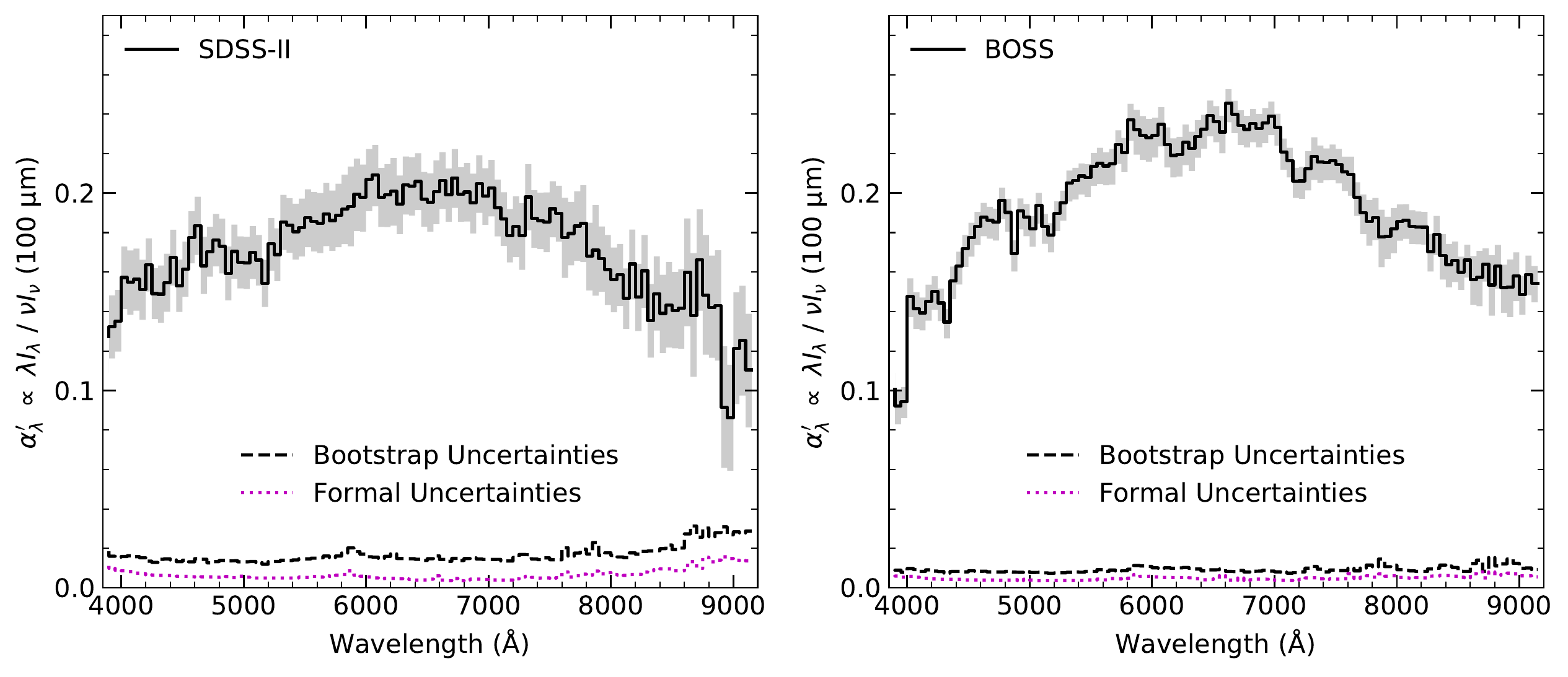}
\caption{DGL correlation spectra for SDSS-II (left) and BOSS (right), both computed using our updated model (Section~\ref{sec:updated_model}). The BOSS spectrum has a similar shape but smaller uncertainties. Small differences in spectral shape are most likely due to increased coverage of the southern Galactic hemisphere by BOSS (Section~\ref{sec:spatial}). The envelopes around the spectra are 68\% confidence intervals found with bootstrapping (Section~\ref{sec:bootstrapping}), and the errors plotted with dashed lines are half of this interval, corresponding to $1\sigma$ limits if the errors are Gaussian. The dotted lines are formal uncertainties based on errors given by the SDSS-II and BOSS pipelines; they underestimate the true uncertainties. The data are binned with a bin width of 50\,\AA{} and nebular emission lines are masked.}
\label{fig:compare_errs}
\end{figure*}

Figure \ref{fig:compare_errs} shows spectra of the DGL computed using the approach described in Section~\ref{sec:updated_model}, with uncertainties from bootstrapping (Section~\ref{sec:bootstrapping}). 
The spectrum in the left panel is based on data from SDSS-II, and the right panel shows the new spectrum using data from BOSS. 

The SDSS-II and BOSS correlation spectra, overall, are very similar. 
However, the BOSS spectrum is slightly more peaked near 6500\,\AA{}, and the optical intensity relative to 100\,$\upmu$m emission appears to be higher for the BOSS spectrum at most wavelengths.
We show in Section~\ref{sec:spatial} that these differences can be explained by sky coverage, and that some of the difference may be due to spatial variation in the uncertain bias factor (Section~\ref{sec:bias_factor}) rather than a true difference in emission.

The BOSS DGL spectrum represents a substantial gain in precision over the spectrum from SDSS-II, largely due to the increased number of optical spectra. Between 5000\,\AA{} and 8000\,\AA, with nebular emission lines masked and the spectra binned into 50\,\AA{} bins, the median signal-to-noise ratio is 24 for BOSS compared to 12 for SDSS-II. This difference is even more apparent near the edges of the SDSS-II wavelength range because BOSS covers a wider range.

Bootstrap uncertainties in the binned spectra are larger than the formal uncertainties calculated from errors given by the SDSS-II and BOSS pipelines. The dominance of bootstrap uncertainties makes the larger sample size and more uniform spatial sampling of BOSS especially important.

\subsection{Updated Model Results}
\label{sec:updated_results}
 
\begin{figure*}
\includegraphics[width=\textwidth]{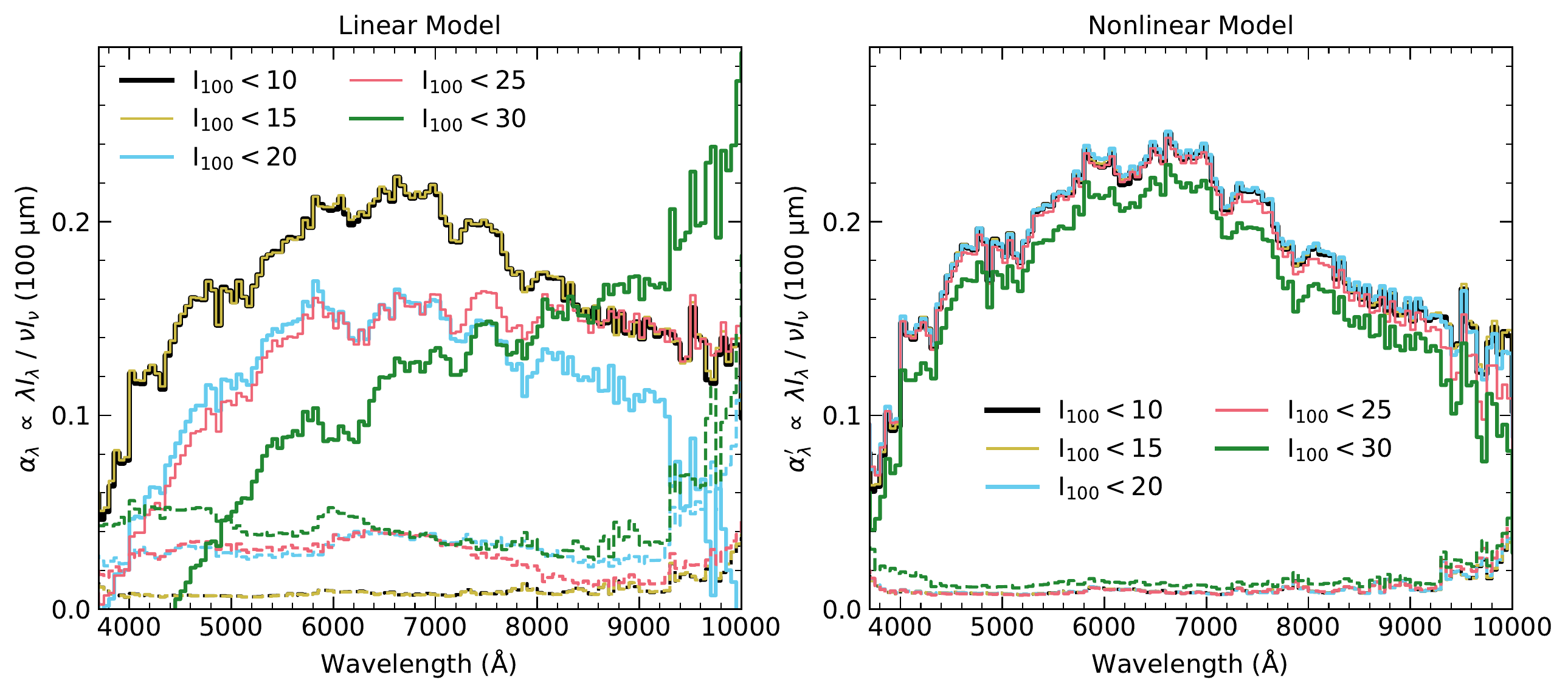}
\caption{Left panel: the linear correlation model of \citetalias{Brandt+Draine_2012} fails at high optical depth due to self-absorption by dust; the DGL spectrum depends on the 100\,$\upmu$m intensity threshold (given in MJy\,sr$^{-1}$) used to mask plates and fibers. Right panel: our updated correlation model for the DGL (Section~\ref{sec:updated_model}) accounts for self-absorption by dust and is valid in regions of modest optical depth. The resulting spectrum is insensitive to the threshold used to mask fibers. We plot $\alpha_\lambda$ in the left panel, but in the right panel we plot $\alpha_\lambda^{\prime}$ (Equation~\eqref{eqn:alpha_prime}), which is more appropriate for comparison. The dashed lines are half of 68\% confidence intervals obtained from bootstrapping (Section~\ref{sec:bootstrapping}). The data are binned with a bin width of 50\,\AA{} and nebular emission lines are masked.}
\label{fig:thresholds}
\end{figure*}

In Section~\ref{sec:updated_model} we describe a new model for the DGL that accounts for dust self-absorption. 
Figure~\ref{fig:thresholds} shows that this updated model does indeed reduce the need to mask regions of high dust optical depth, as it makes our results insensitive to the choice of masking threshold.
The left panel adopts the linear model of Section~\ref{sec:linearmodel} with a range of masking thresholds, while the right panel applies the same thresholds for the nonlinear model of Section~\ref{sec:updated_model}.

For low 100\,$\upmu$m masking thresholds (e.g. 10 and 15~MJy\,sr$^{-1}$), the optical depth is small enough that the linear model is a reasonable approximation; the updated model gives similar results. As expected, the biggest correction is at the blue end of the spectrum where there is more self-absorption. The correction at these wavelengths is $\sim$20\%, with the linear model underestimating the DGL intensity due to its neglect of self-absorption.

As the masking threshold increases, fibers with higher optical depth are included and effects from absorption become more significant. The results of the linear model change drastically, indicating that the model is breaking down. In addition, the newly-included plates have a large impact on the spectrum, increasing the bootstrap uncertainties. On the other hand, the results from the nonlinear model remain virtually identical even with a masking threshold of 25~MJy\,sr$^{-1}$. At this intensity the optical depth for 18\,K dust ranges from $\tau \approx 2$ at 4000\,\AA{} to $\tau \approx 0.5$ at 10,000\,\AA.

\subsection{Spatial Variation}
\label{sec:spatial}

\begin{figure}
\includegraphics[width=\linewidth]{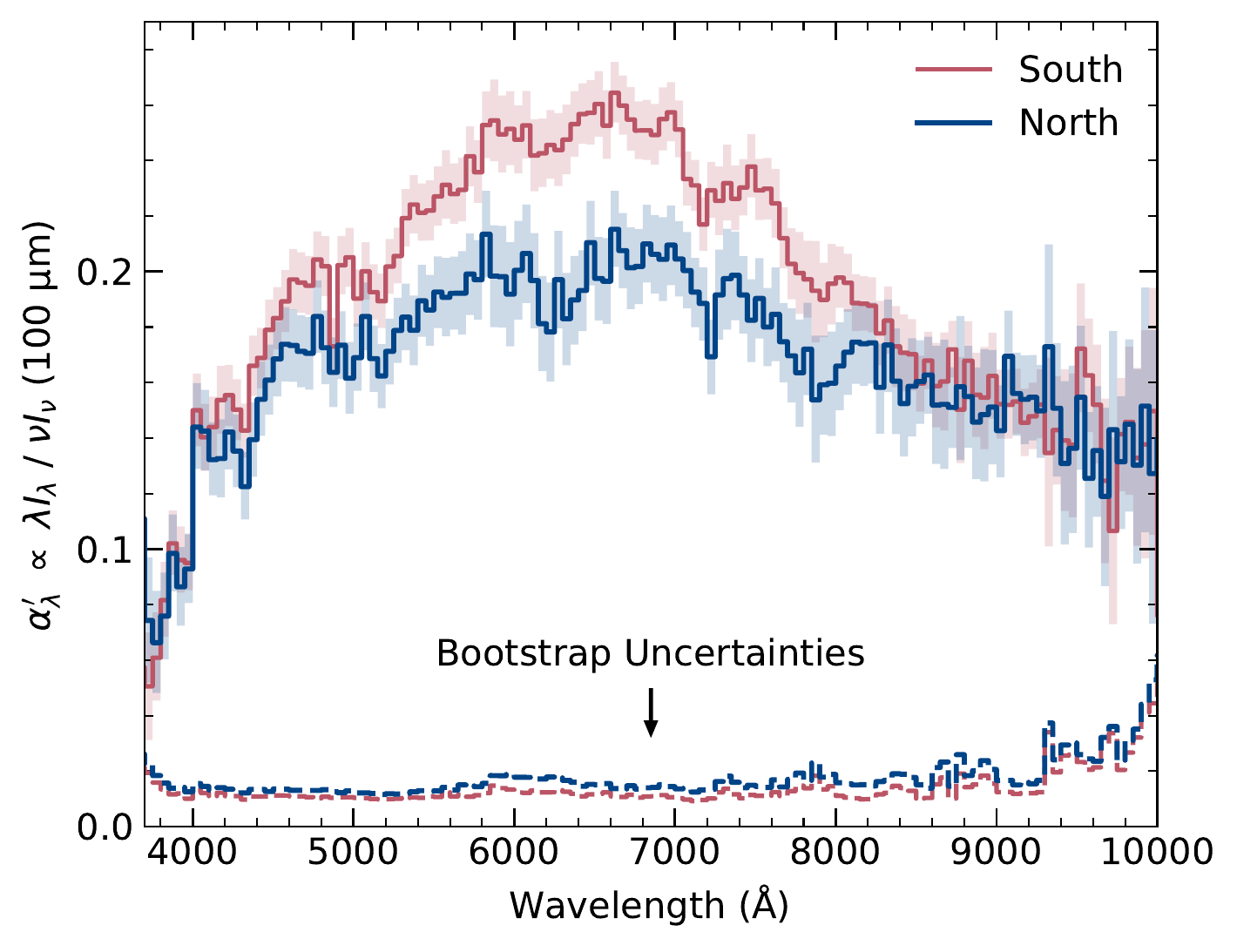}
\caption{DGL correlation spectra for the regions of the BOSS survey in the northern and southern Galactic hemisphere. The two regions ($b > 0$ and $b < 0$) can be clearly distinguished in Figure~\ref{fig:skymaps}. Much of the difference in the spectra seen here could be explained by a difference in the uncertain bias factor (Section~\ref{sec:bias_factor}), but there may also be signs of differences in stellar populations and ERE (Sections~\ref{sec:starlight} and \ref{sec:ere}). The envelopes around the spectra are 68\% confidence intervals found with bootstrapping (Section~\ref{sec:bootstrapping}), and the errors plotted with dashed lines are half of this interval. The data are binned with a bin width of 50\,\AA{} and nebular emission lines are masked.}
\label{fig:north_south}
\end{figure}

The DGL spectrum is expected to vary across the sky due to differences in the interstellar radiation field and/or dust properties. 
In Figure~\ref{fig:north_south} we plot DGL spectra for the footprints of BOSS in the northern ($b > 0$) and southern ($b < 0$) Galactic hemispheres. Subdividing the data further quickly results in spectra with bootstrap uncertainties too large for detecting spatial variations.

This division is a natural choice for comparisons with SDSS-II, which observed a similar region to BOSS in the north but had limited and patchy sky coverage in the south. In fact, the SDSS-II correlation spectrum closely matches the north BOSS spectrum, suggesting that the difference seen in Figure~\ref{fig:compare_errs} between the SDSS-II and BOSS spectra can be attributed to sky coverage.

While the difference between the north and south spectra in Figure~\ref{fig:north_south} initially appears highly significant, we find that much of it can be explained by spatial variation in the bias factor (Section~\ref{sec:bias_factor}). 
\edit1{In Section~\ref{sec:fitting} we adopt bias factors of $C=2.1$ in the north and $C = 1.9$ in the south; when this scaling is applied,} the north and south spectra agree well with each other and with plausible stellar models over \edit1{the range 4200\,\AA{} to 5000\,\AA{}}.
\edit1{F}or wavelengths above 5500\,\AA, there are significant differences in the spectra that cannot be easily resolved with a scaling factor\edit1{, and we explore these} in Sections~\ref{sec:starlight} and \ref{sec:ere}. Possible explanations include differences in stellar populations, dust properties, and ERE.

\edit1{As in \citetalias{Brandt+Draine_2012}, we observe significant variation in the DGL spectrum with Galactic latitude,
but there does not appear to be a clear trend in the latitude dependence.
Several effects could contribute including variation in dust properties and variation in the ISRF.
Especially at high latitudes, where the signal is weaker, there may also be contamination by extragalactic light \citep{2007PASJ...59..205Y} and systematics from an imperfect model. It is difficult to separate these effects without a more detailed model of the galaxy.}

\subsection{Emission Lines} 
\label{sec:lines}

\begin{figure*}
\includegraphics[width=\textwidth]{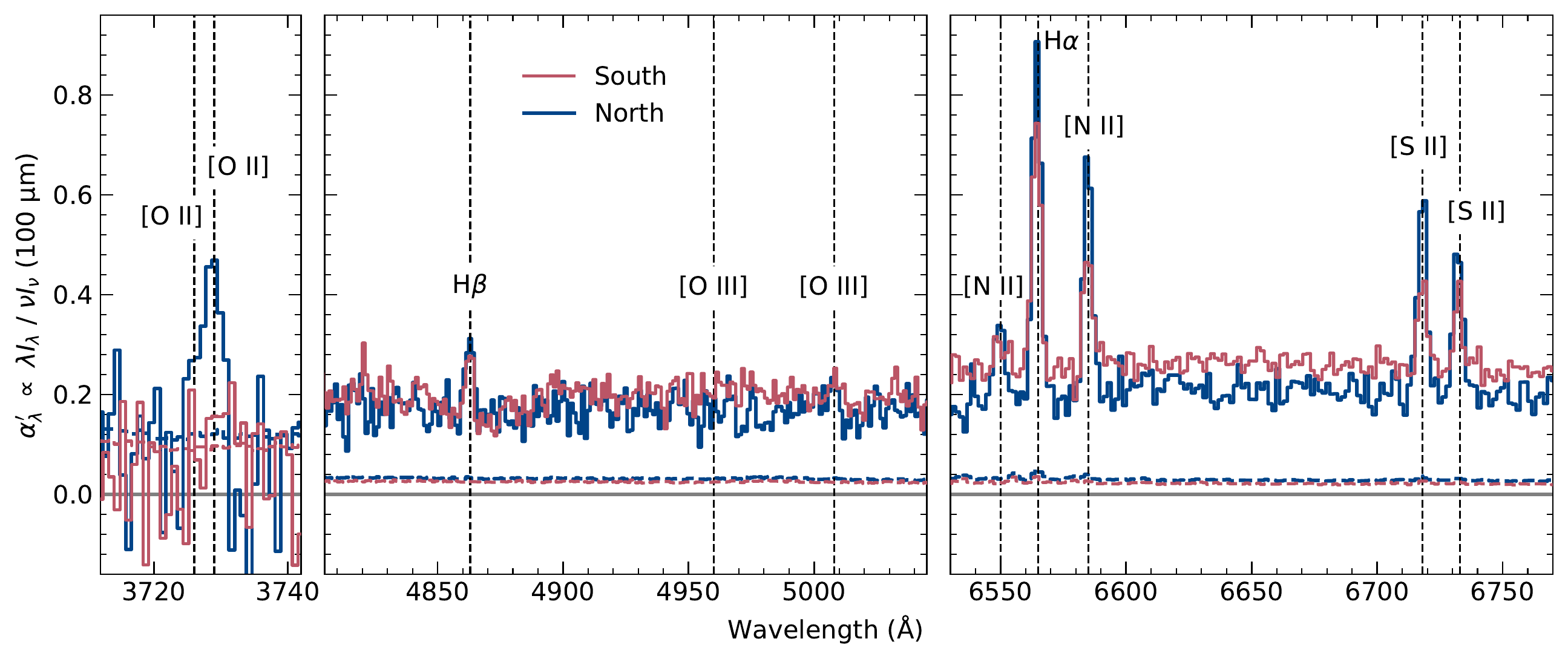}
\caption{Unbinned DGL correlation spectra around prominent nebular emission lines. Spectra are shown for the northern and southern Galactic hemisphere. Left: the [{\sc O\,ii}]\,$\lambda\lambda$3727-30\,\AA{} doublet is visible with BOSS due to the increased wavelength range compared to SDSS-II; note that the horizontal scaling for the first panel differs from the other panels. Equivalent widths are reported in Table~\ref{tab:widths} and are generally larger in the north due to a greater contribution from ISM light. The errors plotted with dashed lines are half of the 68\% confidence intervals found with bootstrapping (Section~\ref{sec:bootstrapping}).}
\label{fig:peaks}
\end{figure*}

Peaks from nebular emission are clearly visible in Figure~\ref{fig:peaks}; they represent light emitted by interstellar gas and scattered by dust grains. We mask these lines in our analysis and in most plots, but the line strengths can be useful in their own right as tracers of ISM properties; they carry signatures of the characteristic densities, temperatures, and ionization states of the nearby ISM.

Table~\ref{tab:widths} reports the equivalent widths of nebular emission lines for the full sky and for the northern and southern Galactic hemisphere. The listed uncertainties are half of the 68\% confidence intervals found with bootstrapping (Section~\ref{sec:bootstrapping}).

We calculate the equivalent widths by fitting Gaussians, assuming a flat continuum and using the maximum likelihood estimator to simultaneously fit the continuum level and height of each peak. We center the fits on the known rest wavelengths of the lines and assume widths corresponding to instrumental broadening from the BOSS spectrograph. This gives a width for the H$\alpha$ line that is in good agreement with the value we find by including the width as an additional parameter in the fit and performing a 1D nonlinear optimization.

The one exception to this is the [{\sc O\,ii}]\,$\lambda\lambda$3727-30\,\AA{} doublet, which is outside the SDSS-II wavelength range but inside the range of BOSS. In principle the relative line strength can serve as an indicator of electron density \citep{Draine_2011a}, however the noise level is too high to provide meaningful constraints. Instead we assume a ratio of [{\sc O\,ii}]$\lambda$3730/[{\sc O\,ii}]$\lambda3727 = 1.5$, corresponding to the limit of low electron density, and fit a sum of two Gaussians. This limit is implied by the values we find for [{\sc S\,ii}]$\lambda$6718/[{\sc S\,ii}]$\lambda$6733, and it is consistent with our expectations for the Warm Ionized Medium \citep{Madsen+Reynolds+Haffner_2006} and for low-density {\sc H\,ii} regions such as the one around the nearest O-type star, $\zeta$~Oph \citep{Maiz-Apellaniz+Walborn+Galue+Wei_2004}.

H$\alpha$ and H$\beta$ are present in absorption in starlight, so our fit underestimates their strength in emission. We correct for this using Equations~(8) and (9) of \citetalias{Brandt+Draine_2012} which are approximate linear relations between the size of the 4000\,\AA{} break and the equivalent widths of the H$\alpha$ and H$\beta$ absorption lines.

The equivalent widths are generally larger in the north, which appears to be due to a greater share of the interstellar radiation field coming from gas emission. The line ratios in the north and south are similar, suggesting similar physical ISM states.

\section{Starlight and Dust} 
\label{sec:starlight}

The interstellar radiation field at optical wavelengths is dominated by starlight, so we expect the DGL to be mainly composed of scattered starlight. Here we confirm this by comparing our measured DGL correlation spectra to model stellar spectra.
We then attempt to determine the properties of the stars that are responsible and to provide some constraints on the properties of dust grains.
Because we are concerned here with starlight, we mask wavelengths where we expect nebular emission lines from interstellar gas, as described in Section~\ref{sec:updated_model}.

\subsection{Models}
\label{sec:models}

The stellar population in the solar neighborhood covers a range of ages, but with a higher proportion of old stars than young stars \citep{Casagrande+Schonrich+Asplund+etal_2011}. Our starlight templates are composite stellar populations from \citet[][hereafter \citetalias{Bruzual+Charlot_2003}]{Bruzual+Charlot_2003}: exponentially declining star formation over 12~Gyr with timescales of 5\,Gyr and 9\,Gyr (``t5e9'' and ``t9e9'') and constant star formation over 6\,Gyr (``cst''). 
We consider linear combinations of these model spectra with metallicity $Z=0.02$.

To account for scattering and extinction by dust, we apply the same radiative transfer model described in Section 5.1 of \citetalias{Brandt+Draine_2012}, summarized here.
We assume an infinite plane-parallel galaxy and a sight line with latitude $b = 40^\circ$. The dust distribution is Gaussian: $\rho_\textrm{dust}=\exp{\left(-z^2/2\sigma^2\right)}$ with $\sigma$ = 250\,pc. We assume a $V$-band dust optical depth of $\tau_V = 0.15 \csc b$.
There are two exponential stellar distributions with scale heights of 300 and 1350\,pc, and the 300\,pc component contains 90\% of the stars.
For scattering we use a Henyey-Greenstein phase function.
The total absorbed power is taken as an estimate of the total power radiated in the IR, which is then converted to the 100\,$\upmu$m bandpass using the \cite{Draine+Li_2007} dust model. The relation is $(\nu I_{\nu})_{100\,\upmu \textrm{m}} \approx 0.52 I_{\textrm{TIR}}$. This allows us to find the ratio of scattered light to 100\,$\upmu$m emission, $\lambda I_{\lambda}$\,/\,$(\nu I_{\nu})_{100\,\upmu \textrm{m}}$, which can be compared to $\alpha_\lambda$ (Equations~\eqref{eqn:alpha_corr} and \eqref{eqn:alpha_prime}).
The implementation of this radiative transfer model in \citetalias{Brandt+Draine_2012} included two errors which we correct. These turned out to be largely compensating errors; our final results are very similar to those of \citetalias{Brandt+Draine_2012}.

We consider dust models from \cite{Zubko_2004} and \cite{Weingartner_2001}, hereafter \citetalias{Zubko_2004} and \citetalias{Weingartner_2001}. Both are ``bare'' models consisting only of PAHs, graphite grains, and silicate grains. Values for albedo, wavelength-dependent cross section, and anisotropy parameter ($g=\cos{\theta}$) are taken from these models.
The \citetalias{Weingartner_2001} model contains more large grains, resulting in a redder spectrum as shown in Figure~3 of \citetalias{Brandt+Draine_2012}.

\subsection{Evidence of Starlight}

We expect the DGL spectrum to be dominated by scattered starlight, and we see convincing evidence that this is the case, both in the spectral features of binned plots and in a more detailed comparison to model spectra. 
This serves as a check that our method is working properly and suggests that we should be able to determine properties of stars and dust grains by comparing to model spectra.
In the right panel of Figure~\ref{fig:compare_errs} (and in other binned plots) there are clear indicators of starlight, including the 4000\,\AA{} break and absorption near 5170\,\AA.

\begin{figure*}
\includegraphics[width=\textwidth]{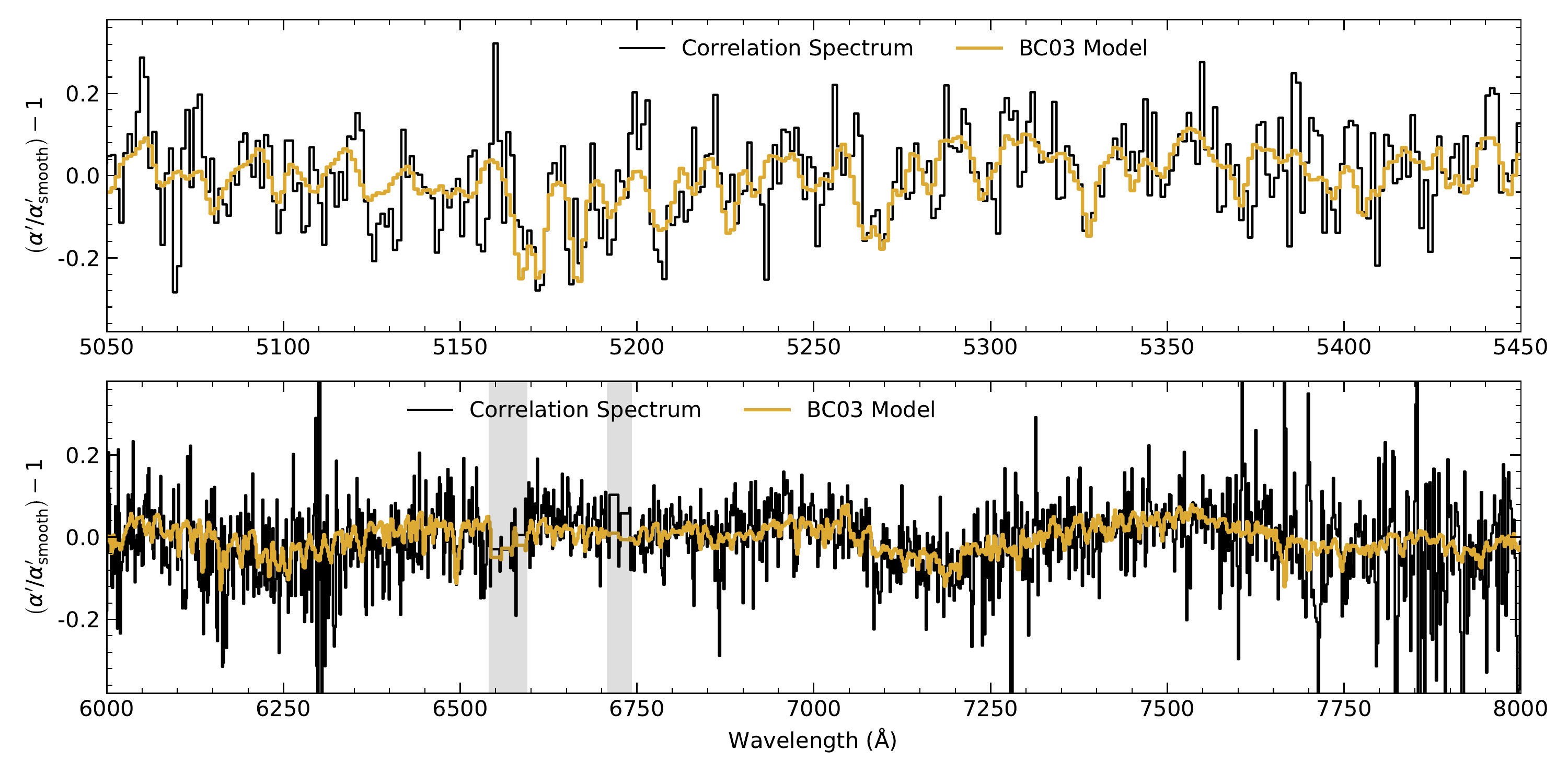}
\caption{Unbinned DGL correlation spectrum plotted together with a \citetalias{Bruzual+Charlot_2003} model stellar spectrum (exponentially declining star formation over 12~Gyr with a timescale of 5~Gyr, $Z=.02$). The upper panel is 400\,\AA{} wide while the lower panel is 2000\,\AA{} wide, and the observed spectrum agrees well with starlight on both scales. Other \citetalias{Bruzual+Charlot_2003} models give similar results. To remove the continuum we divide each spectrum by a polynomial fit. The original model resolution corresponds to a velocity dispersion of $\sigma_v = 70$~km\,s$^{-1}$, and we smooth with a Gaussian kernel corresponding to $\sigma_v = 90$~km\,s$^{-1}$, the value that results in the best fit. The gray shaded regions in the lower plot cover gaps in the data due to emission line masking.}
\label{fig:unbinned_compare}
\end{figure*}

In Figure~\ref{fig:unbinned_compare} we plot the correlation spectrum at the native BOSS resolution alongside a model from \citetalias{Bruzual+Charlot_2003} (exponentially declining star formation over 12~Gyr with a timescale of 5~Gyr, $Z=.02$).
We divide each spectrum by a high-order polynomial to remove the continuum, using least squares to fit the polynomial coefficients. This takes into account the higher uncertainties toward the edges of the wavelength range. While not physically motivated, we find that a 10\textsuperscript{th} order polynomial is sufficient to remove the power from the continuum.

The observations agree with the model of scattered starlight better than with a flat continuum ($\chi^2_\nu$ = 0.60 versus 0.69).
The upper panel of Figure \ref{fig:unbinned_compare} (400\,\AA{} wide) shows that the observed DGL spectrum matches the features of the stellar model quite well, down to the smallest scale resolved by the model. The lower panel (2000\,\AA{} wide) shows that the DGL spectrum also closely follows larger-scale variations in the stellar model.

The resolution of the \citetalias{Bruzual+Charlot_2003} model spectrum corresponds to a velocity dispersion of roughly $\sigma_v = 70$~km\,s$^{-1}$. We smooth with a Gaussian kernel corresponding to $\sigma_v =90$~km\,s$^{-1}$, which is the value that minimizes the $\chi^2$ between the observed spectrum and the model. Adding 70~km\,s$^{-1}$ and 90~km\,s$^{-1}$ in quadrature results in an effective velocity dispersion of $\sigma_v \approx 115$~km\,s$^{-1}$, consistent with the resolution of the BOSS spectrograph and somewhat larger than the velocity dispersion of disk stars \citep{Holmberg_2007}.

\subsection{Fitting Procedure}
\label{sec:fitting}

What types of stars are responsible for the DGL spectrum? We consider linear combinations of three models: exponentially declining star formation over 12~Gyr with timescales of 5~Gyr and 9~Gyr (``t5e9'' and ``t9e9'') and constant star formation over 6~Gyr (``cst''). Although we attempt to quantitatively constrain stellar populations, the results are inconclusive and so we perform a heuristic comparison between the observed spectra and stellar models.

We first try simultaneously fitting several Lick indices which \citetalias{Bruzual+Charlot_2003} found to be sensitive to age and metallicity (Section~4.3 of \citetalias{Bruzual+Charlot_2003}). Unfortunately, some are Balmer series indices which we are unable to use because we cannot accurately separate the contributions from nebular emission.
The fits from all three spectra are similar enough that we are unable to determine a clear best-fit combination.
We also try a fit to the whole spectrum after continuum normalization but are again unable to draw a meaningful conclusion.

\begin{figure*}
\includegraphics[width=\textwidth]{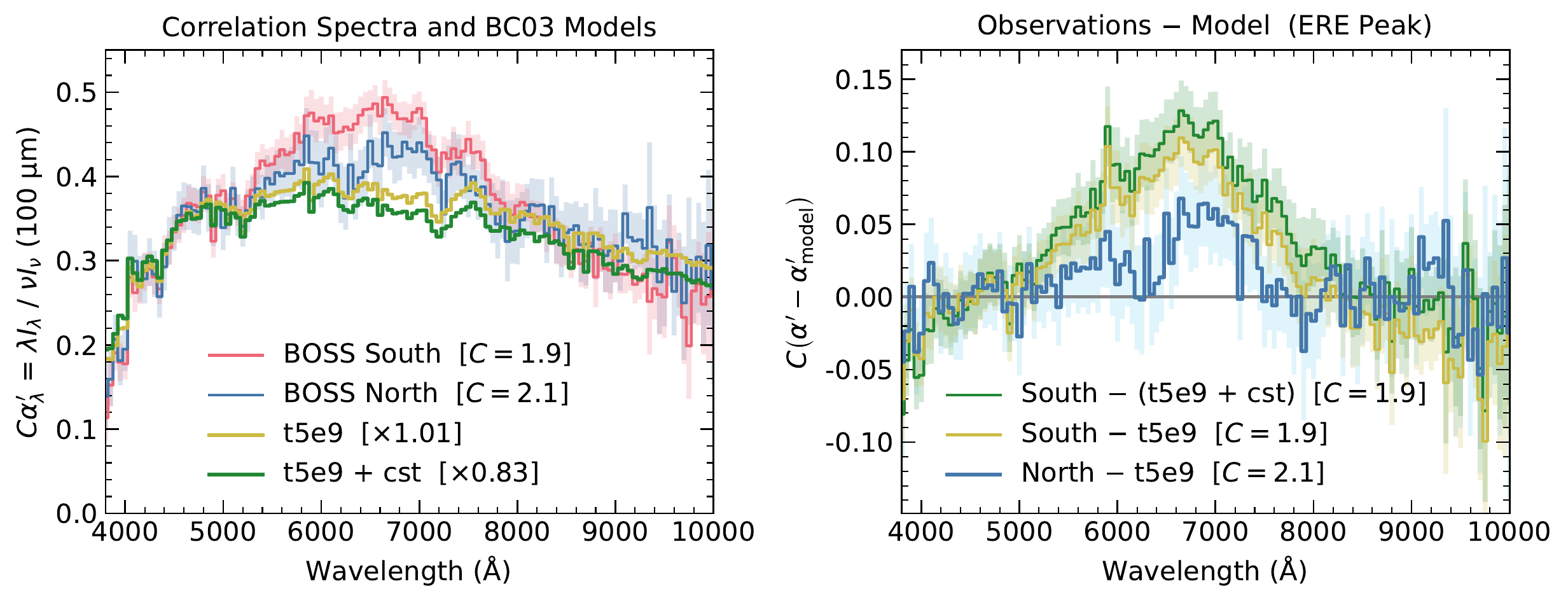}
\caption{Left panel: measured DGL correlation spectra, together with predictions derived from \citetalias{Bruzual+Charlot_2003} starlight models using a simple radiative transfer model (Section~\ref{sec:models}) and the \citetalias{Zubko_2004} dust model. Our choice of starlight models is described in Section~\ref{sec:stellar_pops}. We scale all spectra to a common level over the wavelength range 4200-5000\,\AA{}, which corresponds to applying bias factors (Section~\ref{sec:bias_factor}) of $C=2.1$ in the north and $C=1.9$ in the south. The southern DGL spectrum shows a clear excess over the models at $\sim$6500\,\AA, and we interpret this as ERE, while the excess in the north is less clear. Right panel: we plot the excess directly to visualize the ERE peak. The envelopes around the spectra are 68\% confidence intervals found with bootstrapping (Section~\ref{sec:bootstrapping}). The data are binned with a bin width of 50\,\AA{} and nebular emission lines are masked.}
\label{fig:ere_2plot}
\end{figure*}

Our best constraints, then, come from fitting by eye. The left panel of Figure~\ref{fig:ere_2plot} shows fits to the north and south BOSS spectra, determined in Section~\ref{sec:stellar_pops} assuming \citetalias{Zubko_2004} \edit1{dust.
We} primarily consider wavelengths toward the edges of the range when evaluating fits because of a possible contribution from ERE (Section~\ref{sec:ere}).

\edit1{To estimate} the uncertain bias factor (Section~\ref{sec:bias_factor})\edit1{,} we \edit1{first assume that the predicted spectrum derived from the t5e9 model represents the true level of the DGL spectrum. Then we} scale the \edit1{observed} spectra to match the average level of the \edit1{t5e9 spectrum} over the range 4200-5000\AA{}\edit1{. This suggests bias factors of} $C=2.1$ in the north and $C=1.9$ in the south\edit1{, in agreement with the value of $C = 2.1 \pm 0.4$ adopted by \citetalias{Brandt+Draine_2012}.}

\subsection{Stellar Populations in the North and South}
\label{sec:stellar_pops}

Here we present our best-guess stellar populations based on the fitting method described in the previous section.
For the region covered by BOSS in the northern Galactic hemisphere, we prefer the t5e9 model; it matches the observed DGL spectrum reasonably well when the \citetalias{Zubko_2004} dust model is used, as shown in the left panel of Figure~\ref{fig:ere_2plot}.
The t5e9 model fits better than either t9e9 (more gradually declining star formation) or cst (constant star formation). Both of these models, when combined with \citetalias{Zubko_2004} dust, predict the DGL to be bluer than observed. 

The DGL spectrum in the southern hemisphere appears to be slightly bluer than the spectrum in the northern hemisphere.
The left panel of Figure~\ref{fig:ere_2plot} shows that, with the scaling described in Section~\ref{sec:fitting}, the t5e9 model slightly overpredicts the southern DGL at the reddest wavelengths.
There are several effects that could account for this, including random variation or differences in metallicity, dust properties, or stellar populations. ERE is expected to cause a broad peak, as we discuss in Section~\ref{sec:ere}, but it should not cause a tilt.

A search of Gaia EDR3 \citep{GaiaEDR3} suggests that the southern region of the BOSS survey contains a higher fraction of young stars within 500~pc. As a proxy for young stars, we search for stars that are bluer than $B_p-R_p=0.5$ and have absolute magnitudes brighter than $G=-1$. 
The approximate northern and southern footprint of BOSS each contain 18 such stars within 500~pc of the Sun. However, the northern footprint is much larger, with well over twice as many stars overall.
Therefore a difference in stellar populations is a plausible explanation.

We see in the left panel of Figure~\ref{fig:ere_2plot} that adding constant star formation makes the spectrum bluer and improves agreement with the measured DGL spectrum in the south.
We add enough constant star formation to match the level of the south spectrum at 9000\,\AA. Our preferred model in the south is then a linear combination of the t5e9 and cst models, with $\sim$40\% of the present-day star formation due to the cst population.

\subsection{Dust Models and Systematics}
\label{sec:dust_models}

The synthetic spectra in Figure \ref{fig:ere_2plot} are computed using the \citetalias{Zubko_2004} dust model.
Here we consider the \citetalias{Weingartner_2001} dust model, which includes more large grains and yields a redder scattered spectrum.

The \citetalias{Zubko_2004} model results in a slightly better fit, particularly in the south. The same stellar models with the \citetalias{Weingartner_2001} dust model overpredict the DGL spectrum at red wavelengths.
The level of the south spectrum at 9000\,\AA{} can still be matched by adding more constant star formation, with this population contributing $\sim$60\% of the present-day star formation.
The resulting spectral shape, however, does not agree quite as well. 

Unfortunately, the bootstrap uncertainties and the limited nature of our radiative transfer calculations prevent us from drawing strong conclusions.
There are possible systematics in our model (as defined in Section~\ref{sec:models}), including effects from our choice of model parameters, which could change our dust model preference.
This would have only a minor impact on our conclusions about stellar populations, with both the north and south spectra remaining best modeled with mostly old stars, but varying the spectral shape could tip the balance in favor of the \citetalias{Weingartner_2001} dust model. 

Changing the metallicity from $Z=0.02$ to $Z=0.008$ results in a 9\% difference in the predicted correlation spectrum at 9000\,\AA. Changing the value of $\tau_V$ at $b = 90^\circ$ from 0.15 to 0.05, a plausible value based on Figure~11 of \citetalias{Brandt+Draine_2012}, results in a difference of 22\%. 
These changes both lower the spectrum at red wavelengths (with the scaling described above, where the blue end is held fixed), and they are both comparable to the 15\% difference caused by switching to the \citetalias{Weingartner_2001} dust model.
Other parameter choices have a smaller impact on the spectrum. These include changing the latitude of the sight line from $b=40^\circ$ to $20^\circ$ or $60^\circ$ and removing the thick-disk stellar distribution with a scale height of 1350\,pc.

\section{ERE Detection}
\label{sec:ere}

Our DGL spectra, particularly in the southern Galactic hemisphere, show a broad excess over stellar models that is centered near 6500\,\AA. A wide range of previous works have found a similar excess, known as ERE, in reflection nebulae \citep[e.g.][]{Schmidt+Cohen+Margon+etal_1980,Witt+Schild+Kraiman_1984,1990ApJ...355..182W} and the diffuse ISM (e.g. \citet[][hereafter G98]{Gordon1998DetectionOE}; \citet[][hereafter SG98]{Szomoru+Guhathakurta_1998}). 
The carrier responsible is unknown; see \cite{witt_2020} for a recent review of constraints on ERE models.
Observations of a variety of objects and environments enable comparisons that can help identify the ERE carrier.

The first measurements of ERE were made in reflection nebulae, and due to similarities in dust properties, it was predicted that ERE would also be found in the diffuse ISM of the Milky Way \citep{witt_2020}. 
Since then there have been several detections. Most have relied on broad-band measurements of sky intensity in a small number of bands, including \citetalias{Gordon1998DetectionOE} and \citet{witt_2008}, which looked at the diffuse ISM.
\citetalias{Szomoru+Guhathakurta_1998} made spectroscopic measurements of three high-latitude dust clouds. Our spectroscopy of the diffuse ISM complements these nicely.

\subsection{Our Detection}
\label{sec:our_detection}

Here we describe our detection of ERE.
In the left panel of Figure~\ref{fig:ere_2plot} we plot our DGL spectra for the regions covered by BOSS in the northern and southern Galactic hemisphere, with estimates of the uncertain bias factor (Section~\ref{sec:bias_factor}) applied.
The south spectrum appears slightly more peaked, suggesting the presence of ERE.
In Section~\ref{sec:stellar_pops} we identified various effects that can introduce a tilt in the predicted DGL spectrum: differences in stellar populations, dust properties, and metallicity. However, none of these can reproduce a broad peak.

Next we show that a comparison to models of scattered starlight provides further evidence for ERE.
Unlike in the case of reflection nebulae, we cannot directly measure the spectrum incident on the dust, so we instead make use of \citetalias{Bruzual+Charlot_2003} model spectra. We apply the radiative transfer model from Section~\ref{sec:models} and fit by eye at $\sim$4500\,\AA{} and 9000\,\AA{} as described in Section~\ref{sec:fitting}. 
In the left panel of Figure~\ref{fig:ere_2plot} we plot our preferred model spectra (Section~\ref{sec:stellar_pops}) together with the observed spectra.
The right panel shows the excess in the observed spectra compared to the model predictions; this excess serves as our estimate of ERE.

For the DGL spectrum in the northern BOSS region, there does appear to be some excess at $\sim$6500\,\AA{}, but the noise level is too high to claim a significant detection. 
For the spectrum in the southern region, on the other hand, there is a significant peak in the excess compared to our best-guess model (green curve; a combination of exponential and constant star formation) over the range 5000-8000\,\AA{}. We also plot the peak for only exponential star formation (gold curve), but there is little difference in the intensity or shape of the inferred ERE spectrum.
The higher significance of our ERE detection in the south is partly because the spectrum is more peaked, and partly because of smaller uncertainties.
This spatial difference may explain why \citetalias{Brandt+Draine_2012} did not claim a detection of ERE: SDSS-II sampled mostly from the northern hemisphere.

In Figure~\ref{fig:ere_2plot} we include bootstrap envelopes for the observed spectra, but we neglect model systematics, which also contribute to the uncertainty.
The width and size of the ERE peak depend to some extent on choices we make about scaling, the dust model, and the mix of stellar populations. The model fit is quite good at blue wavelengths (4200-5000\,\AA), but the rest of the systematics we have considered (Sections~\ref{sec:stellar_pops} and \ref{sec:dust_models}) affect the level of the spectrum at red wavelengths. 
Any combination of dust model and stellar population that matches observations at the red and blue end underpredicts the spectrum in the middle, suggesting an ERE peak.
It is possible that ERE could extend to longer wavelengths, in which case we would be underpredicting the ERE by choosing a model that matches the DGL spectrum at $\sim$4500\,\AA{} and 9000\,\AA.

The plots in Figure~\ref{fig:ere_2plot} use the \citetalias{Zubko_2004} dust model. If the \citetalias{Weingartner_2001} model is used instead, the t5e9 spectrum predicts a smaller ERE peak but also overpredicts the DGL intensity at red wavelengths. When constant star formation is added to match the level of observations at the red end of the spectrum, the ERE estimate for \citetalias{Weingartner_2001} dust is almost the same as for \citetalias{Zubko_2004} dust.

Previous studies have suggested that the ERE carrier is excited primarily by UV photons \citep[e.g.][]{1999A&A...348..990D, 2006ApJ...636..303W, lai_2017}. 
Our results are consistent with this, since we find stronger evidence for ERE in the southern region of the BOSS survey. As discussed in Section~\ref{sec:stellar_pops}, Gaia EDR3 identifies a higher proportion of luminous (absolute $G < -1$\,mag), relatively blue ($B_P - R_P < 0.5$\,mag) stars within the southern BOSS footprint.

\subsection{Numerical Estimates}
\label{sec:num_est}

In this section we quantify our ERE detection by subtracting the predictions of our radiative transfer model from the observed DGL spectra, using our preferred stellar populations and dust model.

Our correlation spectra are biased low (Section~\ref{sec:bias_factor}), and for the purpose of these calculations we multiply by factors of $C=2.1$ in the north and $C=1.9$ in the south (\edit1{Section~\ref{sec:fitting}}) to correct for this. These factors are important for our estimate of the integrated ERE intensity, but they do not affect the significance of our detection or the ratio of ERE intensity to the intensity of scattered starlight. \edit1{}\citetalias{Brandt+Draine_2012} estimated a 20\% uncertainty in their adopted bias factor of $C=2.1$, and we expect this
to be one of the dominant sources of uncertainty except where the bias factor cancels. Bootstrap uncertainties are comparable: we estimate $\sim$15\% in the south for the integrated ERE intensity.

Following the procedure of \cite{1990ApJ...355..182W}, we divide the peak into quartiles of equal integrated intensity, then report the 2\textsuperscript{nd} quartile as the peak and the difference between the 1\textsuperscript{st} and 3\textsuperscript{rd} quartiles as the width. We find a peak of 6500\,\AA{} and a width of 1100\,\AA. 
To get the integrated ERE intensity in units of flux per solid angle, we rearrange Equation~\eqref{eqn:alpha_corr} and multiply by the bias factor $C$ to find that the optical intensity is given by
\begin{equation}
\label{eqn:ere_integral}
I_{\lambda} = \frac{C\alpha_\lambda \left(\nu I_{\nu}\right)_{100\,\upmu \textrm{m}}\beta_\lambda}{\lambda},
\end{equation}
and for $I_{100\,\upmu \textrm{m}}$ we take a mean over the unmasked BOSS sky fibers. 
Integrating the excess in $I_{\lambda}$ over the prediction of our preferred \citetalias{Bruzual+Charlot_2003} model gives $I_{\textrm{ERE}} \approx 0.28 \times 10^{-5}$\,erg\,s$^{-1}$\,cm$^{-2}$\,sr$^{-1}$ in the south (with $\langle I_{100\,\upmu \textrm{m}} \rangle$ = 2.6\,MJy\,sr$^{-1}$). 
In the north we find $I_{\textrm{ERE}} \approx 0.04 \times 10^{-5}$\,erg\,s$^{-1}$\,cm$^{-2}$\,sr$^{-1}$ (with $\langle I_{100\,\upmu \textrm{m}} \rangle$ = 1.4\,MJy\,sr$^{-1}$), although we reiterate that there is not a clear detection in the north.

Next we compute the ratio of integrated ERE intensity to the intensity of scattered starlight over the range 5000-8000\,\AA, roughly corresponding to the photometric $R$ band. This range is chosen because previous works, both photometric and spectroscopic, have reported a similar quantity.
We obtain $I_{\textrm{ERE}} / I_{\textrm{sca}} \approx 0.20$ for the south and 0.06 for the north.
We also compute a ratio of $I_{\textrm{ERE}}$ to total infrared (TIR) intensity, converting from $(\nu I_{\nu})_{100\,\upmu \textrm{m}}$ to $I_{\textrm{TIR}}$ using the \cite{Draine+Li_2007} dust model; \citetalias{Brandt+Draine_2012} estimates a $\sim$10\% uncertainty in this conversion.
The result is $I_{\textrm{ERE}} / I_{\textrm{TIR}} \approx 0.018$ in the south and 0.005 in the north, suggesting that $\sim$1\% of the total power incident on the grains is radiated as luminescence.

Calculating the ratio of $I_{\textrm{ERE}}$ to the UV power incident on the dust allows an estimate of the energy conversion efficiency. We take $\gamma = I_{\textrm{UV}}/I_{\textrm{TIR}}$, roughly equal to the ratio of UV to total incident power.
If ERE is excited by photons more energetic than 6\,eV or 8\,eV, we find using our best-guess \citetalias{Bruzual+Charlot_2003} model that $\gamma \approx 0.56$ or 0.46 in the south, corresponding to 3\% or 4\% of the incident UV power converted to ERE.

\subsection{Comparisons}

Our DGL spectrum provides a measurement of ERE from the diffuse ISM, averaged over large regions of the sky. In this section we compare our ERE constraint with other measurements over smaller spatial regions.

A measurement of ERE requires subtracting the contribution of scattered light, and usually a variety of other backgrounds including airglow and zodiacal light \citepalias[e.g.][]{Szomoru+Guhathakurta_1998}.
\citetalias{Gordon1998DetectionOE} avoided airglow and zodiacal light by using measurements from spacecraft outside the zodiacal dust cloud, but they still had to remove light from stars and galaxies.
Our method has the advantage that we only need to remove the contribution of scattered starlight as described in Section~\ref{sec:our_detection}.
Airglow does not correlate with 100\,$\upmu$m intensity, \edit1{although the SDSS spectra have been sky subtracted, which further helps to isolate the DGL.
Z}odiacal light has been explicitly removed from the \citetalias{Schlegel+Finkbeiner+Davis_1998} and IRIS dust maps. 
\edit1{It is possible that residuals from the zodiacal light removal may be large at high latitudes \citep{2021ApJ...906...77L}, but we find that masking data within $10^\circ$ of the ecliptic has a negligible effect on the DGL spectrum.}
The \edit1{optical} spectra we use have also been placed in regions of the sky without detected optical sources.

The moderate resolution and high signal-to-noise ratio of our DGL spectrum enable a relatively clean detection of the ERE. 
Our peak wavelength of $\sim$6500\,\AA{} is higher than the value of 6000\,\AA{} from \citetalias{Szomoru+Guhathakurta_1998}, but still consistent with the observed trend of ERE peak wavelength increasing with radiation field density \citep{Smith_2002}.

A comparison of our integrated ERE intensity with literature results is difficult.
The bootstrap uncertainty is $\sim$15\% in the south, and we multiply by a bias factor with an uncertainty of $\sim$20\% (Section~\ref{sec:bias_factor}; we adopt $C=2.1$ in the north and $C=1.9$ in the south).
In addition, our method determines a spatially averaged ratio of ERE intensity to 100\,$\upmu$m emission, leaving open the possibility of spatial variation in this quantity. 
With those caveats, the value of $I_{\textrm{ERE}} \approx 1.2\times10^{-5}$\,erg\,s$^{-1}$\,cm$^{-2}$\,sr$^{-1}$ from \citetalias{Szomoru+Guhathakurta_1998} is larger than our inferred value for the southern BOSS footprint by a factor of four, and much larger than our value for the northern footprint.

If we instead consider the ratios $I_{\textrm{ERE}} / I_{\textrm{sca}}$ and $I_{\textrm{ERE}} / I_{\textrm{TIR}}$, as defined in Section~\ref{sec:num_est}, we can avoid some of the problems mentioned above. There should be less spatial variation in these quantities because ERE, optical intensity, and infrared intensity all depend on dust column density. In the case of $I_{\textrm{ERE}} / I_{\textrm{sca}}$, the uncertain bias factor cancels. This ratio is known to depend on scattering geometry \citepalias{Gordon1998DetectionOE}, which can complicate interpretation, but the values here are all from observations taken primarily at Galactic latitudes of $b \gtrsim 25^\circ$.

Our value of $I_{\textrm{ERE}} / I_{\textrm{sca}} \approx 0.20$ is smaller than the value of $\sim$0.3 from \citetalias{Szomoru+Guhathakurta_1998}, although it falls in the range 0.05--2 reported by \citetalias{Gordon1998DetectionOE} who had much better spatial resolution.
\citetalias{Gordon1998DetectionOE} estimates that $I_{\textrm{ERE}} / I_{\textrm{TIR}} \approx 0.03$. For the filaments observed by \citetalias{Szomoru+Guhathakurta_1998}, the 100\,$\upmu$m intensities they report also imply $I_{\textrm{ERE}} / I_{\textrm{TIR}} \approx 0.03$, although they note that zodiacal light has not been subtracted which could increase this fraction. Our determinations of $I_{\textrm{ERE}} / I_{\textrm{TIR}} \approx 0.018$ in the south and 0.005 in the north are smaller than the values found by \citetalias{Gordon1998DetectionOE} and \citetalias{Szomoru+Guhathakurta_1998} unless the correction for bias in determining our spectrum is as large as $C=3$.

\subsection{Blue Luminescence}
\label{sec:bl}

Our DGL spectrum can constrain Blue Luminescence (BL), an emission process that occurs at blue wavelengths \citep{2005ApJ...633..262V}. This constraint exploits the presence of stellar features in the DGL. The size of the 4000\,\AA{} break, a prominent feature of old stellar populations, should be preserved by scattering but diluted if there is any luminescence.

\begin{figure}
\includegraphics[width=\linewidth]{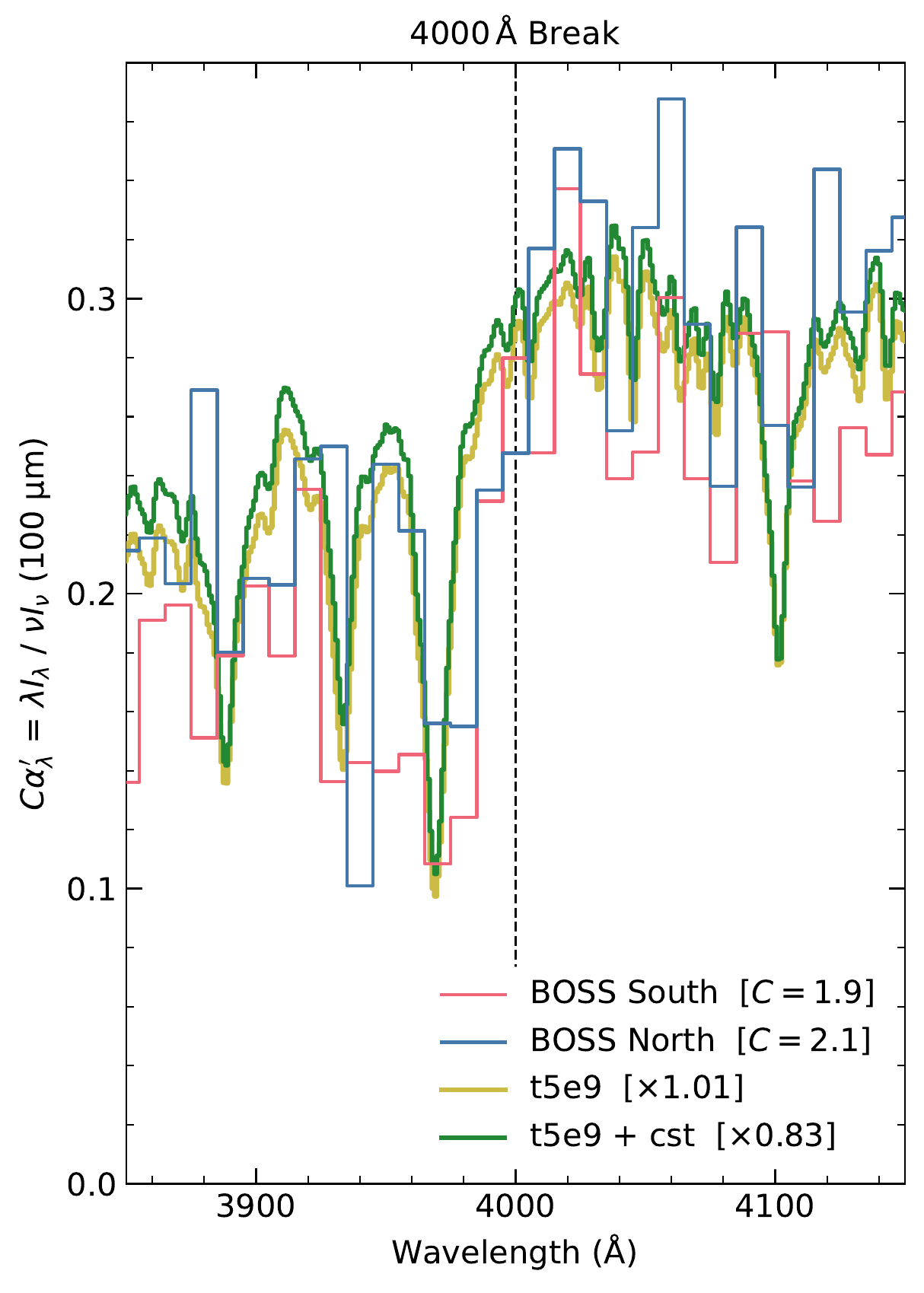}
\caption{Measured DGL correlation spectra around the 4000\,\AA{} break, plotted together with predictions derived from \citetalias{Brandt+Draine_2012} starlight models using a simple radiative transfer model (Section~\ref{sec:models}) and the \citetalias{Zubko_2004} dust model. Our choice of starlight models is described in Section~\ref{sec:stellar_pops}. The wavelength range 3850-4150\,\AA{} is used to quantify the break in Section~\ref{sec:bl}; the break is stronger than expected based on these models. We scale all spectra to a common level over the wavelength range 4200-5000\,\AA, which corresponds to applying bias factors (Section~\ref{sec:bias_factor}) of $C=2.1$ in the north and $C=1.9$ in the south. The observations are binned with a bin width of 10\,\AA{}, while the model spectra are not binned.}
\label{fig:4000}
\end{figure}

In Figure~\ref{fig:4000} we plot the 4000\,\AA{} break for the observed DGL spectra in the north and south and for stellar models; high noise levels and spectral features make visual comparisons of the strength of the break difficult.
We therefore define $\delta_{4000}$ as the ratio of $I_\lambda$ just below the break (3850 to 4000\,\AA) to $I_\lambda$ just above the break (4000 to 4150\,\AA), finding $\delta_{4000} = 0.73 \pm 0.04$ in the south, $0.71 \pm 0.05$ in the north, and $0.72 \pm 0.03$ for the full sky.

The t5e9 \citetalias{Bruzual+Charlot_2003} spectrum and \citetalias{Zubko_2004} dust model predict $\delta_{4000} = 0.80$ for the scattered spectrum; Table~\ref{tab:deltas} reports values of $\delta_{4000}$ for other model combinations.
Thus the observed 4000\,\AA{} break is stronger than expected, even compared to old stellar population models. This is evidence against significant luminescence at 4000\,\AA{}.
If a stellar model correctly describes the 4000\,\AA{} break and the contribution from BL is constant in $I_\lambda$, then the fraction $f_{\textrm{BL}}$ contributed by BL to the total dust-correlated radiation at 4000-4150\,\AA{} is given by
\begin{equation}
f_{\textrm{BL}} = \frac{\delta_{4000} [\textrm{observed}] - \delta_{4000} [\textrm{model}]}{1 - \delta_{4000} [\textrm{model}]}.
\end{equation}
With the full-sky data we use bootstrapping to find a 99.7\% confidence interval for $\delta_{4000}$ of [0.62, 0.81], and from there we can place a 3$\sigma$ upper limit of $f_{\textrm{BL}} = 0.07$, assuming the t5e9 stellar model and \citetalias{Zubko_2004} dust.
If the stellar population contains more young stars, this bound decreases.

\begin{deluxetable}{ccc}
\tablewidth{0pt}
\label{tab:deltas}
\tablecaption{4000\,\AA{} Break for Several Models}
\tablehead{
Stellar Model\tablenotemark{a} &
Dust Model &
$\delta_{4000}$
}
\startdata
t5e9 & \citetalias{Zubko_2004} & 0.80 \\
t5e9 + cst\tablenotemark{b} & \citetalias{Zubko_2004} & 0.82 \\
t5e9 & \citetalias{Weingartner_2001} & 0.80 \\
t5e9 + cst\tablenotemark{c} & \citetalias{Weingartner_2001} & 0.83 \\
\enddata
\tablenotetext{a}{Our best-fit models described in Section~\ref{sec:starlight}}
\tablenotetext{b}{Contributes $\sim$40\% of star formation}
\tablenotetext{c}{Contributes $\sim$60\% of star formation}
\end{deluxetable}

\section{Conclusion}
\label{sec:conclusion}

In this paper we have presented a new measurement of the spectrum of the DGL. This work represents a substantial improvement over the DGL spectrum derived by \citetalias{Brandt+Draine_2012} using a similar approach.
BOSS provides more data with better and more uniform sky coverage, resulting in a spectrum with lower uncertainty, and our model is more robust, applicable in optically thick regions with moderate self-absorption.

The DGL spectrum appears to be largely consistent with a simple radiative transfer model for scattering by dust of starlight from the local stellar population. It reproduces detailed spectral features of starlight, allowing us to infer properties of stellar populations and dust grains.
Comparing results from the regions covered by BOSS in the northern and southern Galactic hemisphere, there is a possible difference at the red end of the spectrum, consistent with a higher proportion of young stars in the southern region.
It is difficult to constrain the dust model, as stronger scattering at red wavelengths may be compensated by a younger, bluer stellar population. With this caveat, we have a slight preference for the \citetalias{Zubko_2004} dust model over the \citetalias{Weingartner_2001} model. The former has fewer large grains and weaker scattering at the red end of the spectrum.

Comparisons to model spectra also reveal a broad excess in the DGL spectrum centered near 6500\,\AA, especially in the southern hemisphere. 
A natural explanation appears to be ERE, supporting previous measurements that have detected ERE in the diffuse ISM.
ERE is believed to result from luminescence of dust grains excited by UV photons, and our stronger detection of ERE in the south is consistent with this; Gaia detects a higher fraction of luminous young stars within 500\,pc in the southern BOSS footprint relative to the northern footprint.
The observed 4000\,\AA{} break is consistent with scattered starlight, with no evidence for BL by dust; we obtain an upper limit of $\sim$7\% on the fraction of dust-correlated light at 4000\,\AA{} contributed by BL.

The large number of sky spectra and the extensive, relatively uniform sky coverage of BOSS provide a powerful new probe of the DGL.
Future analysis could improve the DGL spectrum even further by adding data from surveys including eBOSS \citep{2016AJ....151...44D}, the continuation of BOSS, or DESI \citep{desicollaboration2016}, which will collect an unprecedented number of optical sky spectra. The increase in data should improve spatial resolution, allowing DGL spectra to be computed with acceptable signal-to-noise ratios over smaller patches of sky. In addition, existing and future surveys may allow for dust maps with higher resolution and higher fidelity. These maps would provide better templates for correlating optical intensity, and they could further increase the signal-to-noise ratio and decrease the bias in our results.

Another potential avenue for improving this analysis is the use of {\sc H\,i} maps as a tracer of Galactic dust. At the high Galactic latitudes central to the present analysis, {\sc H\,i} closely traces the dust column density without complications induced by variations in the dust temperature \citep{2017ApJ...846...38L}. The use of spectroscopic {\sc H\,i} data from the HI4PI \citep{2016A&A...594A.116H} and GALFA-HI \citep{2018ApJS..234....2P} surveys may improve upon our use of the IRAS 100\,$\upmu$m map as a dust template.

Future probes of the DGL with increasing precision and better spatial resolution could map the interstellar radiation field, dust properties, and dust luminescence across the local universe.

\acknowledgements{T.D.B.~gratefully acknowledges support from the Alfred P.~Sloan Foundation. Funding for SDSS and SDSS-II has been provided by the Alfred P. Sloan Foundation, the Participating Institutions, the National Science Foundation, the U.S. Department of Energy, the National Aeronautics and Space Administration, the Japanese Monbukagakusho, the Max Planck Society, and the Higher Education Funding Council for England. The SDSS web site is \url{http://www.sdss.org/}. Funding for SDSS-III has been provided by the Alfred P. Sloan Foundation, the Participating Institutions, the National Science Foundation, and the U.S. Department of Energy Office of Science. The SDSS-III web site is \url{http://www.sdss3.org/}.
This work has made use of data from the European Space Agency (ESA) mission Gaia (\url{https://www.cosmos.esa.int/gaia}), processed by the Gaia Data Processing and Analysis Consortium (DPAC, \url{https://www.cosmos.esa.int/web/gaia/dpac/consortium}). Funding for the DPAC has been provided by national institutions, in particular the institutions participating in the Gaia Multilateral Agreement.}
\\
\software{astropy \citep{2013A&A...558A..33A, 2018AJ....156..123A}, numpy \citep{numpy}, scipy \citep{scipy}, matplotlib \citep{matplotlib}}

\bibliography{main}

\end{document}